\documentclass[tightenlines,aps,twocolumn,nofootinbib,superscriptaddress]{revtex4}

\usepackage{float}
\usepackage{amsmath,amsfonts,graphicx,bm}

\newcommand{\be}{\begin{equation}}
\newcommand{\ee}{\end{equation}}
\newcommand{\ba}{\begin{eqnarray}}
\newcommand{\ea}{\end{eqnarray}}

\begin{document}

\begin{titlepage}

\begin{flushright}
\vbox{
\begin{tabular}{l}\\
 UH-511-1125-08
\end{tabular}
}
\end{flushright}

\title{
Next-to-leading order QCD corrections to $t\bar{t}Z$ production at the LHC
}

\author{Achilleas Lazopoulos} 
\affiliation{Department of Physics and Astronomy,
          University of Hawaii,\\ 2505 Correa Rd. Honolulu, HI 96822}  
\author{Thomas McElmurry}
\affiliation{Department of Physics, University of Wisconsin, Madison, WI  53706}
\author{Kirill Melnikov}
\affiliation{Department of Physics and Astronomy,
          University of Hawaii,\\ 2505 Correa Rd. Honolulu, HI 96822} 
\author{Frank Petriello}
\affiliation{Department of Physics, University of Wisconsin, Madison, WI  53706}


\begin{abstract}

We present a calculation of the full next-to-leading order QCD corrections to the scattering process $pp\to t\bar{t}Z$.  This channel 
will be used to measure the $t\bar{t}Z$ electroweak couplings at the Large Hadron Collider.  These couplings cannot be directly measured in current 
experiments.  To obtain these results, we utilize 
a novel, completely numerical approach to performing higher order QCD calculations.  We find that for reasonable values of the 
renormalization and factorization scales the QCD corrections increase 
the leading order result by 35\%, while reducing the theoretical error arising from uncalculated higher order corrections 
to approximately $\pm 10\%$.

\end{abstract}

\maketitle

\end{titlepage}

Probing the properties of the top quark is an urgent priority of the particle physics experimental program.  The large 
mass of the top quark, more than an order of magnitude larger than any other fermion, suggests that it is intimately 
connected to the mechanism of electroweak symmetry breaking and to the generation of hierarchies in the flavor sector of 
the Standard Model (SM).  Although it was discovered more than a decade ago at the Fermilab Tevatron~\cite{topd}, relatively little 
is known about the top quark.  The production process $p\bar{p} \to t\bar{t}$ at the Tevatron proceeds almost entirely through 
gluon exchange, and is sensitive only to the top quark mass and $SU(3)_C$ representation.  The top quark decays primarily 
through $t \to Wb$, leaving only the helicity of the final state $W$ to be measured.  

Details regarding the electroweak properties of the top quark are unknown.  This information requires measurement of the 
$t\bar{t}\gamma$, $t\bar{t}Z$, and top quark Yukawa couplings.  These couplings probe many new physics effects; they are 
sensitive to mixing with additional $Z'$ gauge bosons and new heavy fermions, and can provide access to extended scalar 
sectors.  Although inaccessible at the Tevatron, these interactions can be studied through the radiative processes 
$pp \to t\bar{t}\gamma,Z,H$ at the Large Hadron Collider (LHC).  Measuring deviations from SM predictions can provide 
further information regarding new states discovered at the LHC, or can provide evidence of new physics too heavy 
to produce directly.

The cross sections for the scattering channels $pp \to t\bar{t}\gamma,Z,H$ scale as $\sigma_{\rm LO} \sim \alpha_s^2$ at 
leading order in the QCD perturbative expansion, indicating that there is significant theoretical uncertainty in 
predicting their rates.  It is important to quantify the effect of this error on the measurement of the top quark 
couplings and determine whether a next-to-leading order QCD calculation is needed to improve the theoretical prediction.  A 
procedure for measuring the $t\bar{t}\gamma,Z$ couplings was discussed in~\cite{ulitop}.  This study showed that the 
uncertainties in the signal cross sections hinder the extraction of anomalous top quark couplings.  The 
theoretical error coming from higher order QCD corrections is the limiting factor in measuring $t\bar{t}Z$ couplings at the LHC.

In this Letter we present a full calculation of the next-to-leading order (NLO) QCD corrections to the 
$pp \to t\bar{t}Z$ scattering channel to facilitate accurate measurements of top quark properties at the LHC.  Computations 
of NLO QCD corrections to processes with five or more external particles are notoriously difficult, and 
require overcoming many technical challenges.  Significant community effort has been invested in devising efficient 
calculational algorithms for one-loop calculations~\cite{refs1}, leading
to a host of new results of phenomenological relevance for the LHC~\cite{refs2}.  

Recently, a completely numerical and highly flexible 
approach to NLO and NNLO calculations was suggested 
\cite{lmp,Lazopoulos:2007bv,babis1,babis2}.
It applies 
sector decomposition~\cite{Hepp:1966eg} and contour deformation~\cite{soper} to Feynman 
parametric loop integral representations.  It avoids many pitfalls present in traditional approaches to calculations 
of QCD corrections to multi-particle production processes.  The calculation of the QCD corrections to  
$pp \to t\bar{t}Z$ described here is the first time these
techniques have been applied to a complete scattering channel with multiple mass scales.  
This mode contains the full spectrum of difficulties present in the most complex $2 \to 3$ processes.  
Several five-point topologies with two, three and four internal massive propagators appear.  In this Letter we outline the 
technical details of our calculation and present phenomenological results relevant for LHC analyses.  Preliminary results for 
the $gg$ initiated partonic channel were presented in~\cite{Lazopoulos:2007bv}.

We consider the process $p(P_1) + p(P_2) \to t \bar t Z$. The 
factorization theorems for hard scattering processes in QCD
allow us to write 
\begin{equation}
{\rm d} \sigma = 
\sum_{ij} \int \limits_{0}^{1} {\rm d}x_1 {\rm d}x_2
f^{1}_i(x_1) f^{2}_j (x_2) {\rm d} \sigma_{ij \to t \bar t Z}(x_1 x_2 S),
\label{eq1}
\end{equation}
where $f_{i,j}^{(1,2)}(x_{1,2})$ are the parton densities that give 
the probability to find parton $i(j)$ in the proton $1(2)$ with momentum 
$p_{i(j)} = x_{1(2)} P_{1(2)}$. The center-of-mass energy squared of the 
proton-proton collision is introduced in Eq.~(\ref{eq1}), 
$S = 2P_1\cdot P_2$.  

At leading order in the $\alpha_s$ expansion, 
both $gg$ and $q\bar{q}$ initiated partonic processes contribute.  
The computation of the leading order 
cross section is straightforward. We use QGRAF~\cite{qgraf}  
to generate the relevant 
Feynman diagrams and then MAPLE and FORM \cite{form} 
to manipulate this output.
Throughout the paper we set $m_t = 170.9~{\rm GeV}$, $m_Z = 91.19~{\rm GeV}$, and
$m_W = 80.45~{\rm GeV}$.  For the coupling of the $Z$ boson to quarks, 
we employ 
\begin{equation}
Zqq:~~~i\sqrt \frac{8 m_W^2 G_F}{\sqrt{2} \cos^2 \theta_W} \left ( g^q_v + g^q_a \gamma_5 \right ),
\end{equation}
where $\displaystyle g^q_v = \frac{T_3^q}{2} - Q_q \sin^2 \theta_W,~
g^q_a = -\frac{T_3^q}{2}$,
$\sin^2 \theta_W = 1-m_W^2/m_Z^2 = 0.2215$ is the sine squared 
of the electroweak mixing angle, $T_3^q$ is the weak isospin of the 
quark $q$, $Q_q$ is the electric charge of the quark $q$ in units 
of the proton charge and $G_F$ is the Fermi constant.  Numerical results for the leading order cross section are presented below.

At next-to-leading order, the $qg$ channel additionally contributes.  Several distinct contributions to the NLO cross section 
can be identified:
\begin{itemize}
  \item the one-loop virtual corrections to the leading order processes $gg,q\bar{q} \to t\bar{t}Z$;
  \item the real emission corrections $gg \to t\bar{t}Zg$, $q\bar{q} \to t\bar{t}Zg$ and $qg \to t\bar{t}Zq$;
  \item the renormalization of the leading order cross section.
\end{itemize}
Both the virtual and real emission corrections are separately divergent.  They must be combined to remove divergences 
arising from soft gluon emission.  Ultraviolet divergences are absorbed into the definitions of the coupling constant, 
the top quark mass, and the top quark wavefunction.  Singularities associated with initial state collinear gluon emission 
are absorbed into the definition of the parton distribution functions.  We renormalize the top quark mass 
and wavefunction on shell.  We employ the ${\overline {\rm MS}}$ scheme for the parton distribution functions, and for the QCD coupling 
constant $\alpha_s$ for the five light quarks.  The ultraviolet divergence coming from the top quark is removed through a 
zero-momentum subtraction.  These renormalizations are performed at certain momentum scales.  If computed to all orders in 
perturbation theory the cross section would be independent of this scale choice; the residual dependence of the NLO result on the scale 
provides an estimate of the uncalculated higher order corrections.

We utilize dimensional regularization to control the singularities of the virtual and radiative corrections at intermediate 
stages.  The radiative corrections are organized using the two-cutoff slicing method~\cite{twocut}.  To compute the one-loop 
virtual corrections we employ the strategy we introduced in~\cite{lmp,Lazopoulos:2007bv}.  For each  
one-loop diagram interfered with the full Born amplitude, we analytically perform the integration  
over the loop momentum and arrive at a Feynman parametric representation.  While the ultraviolet 
divergences factorize 
after the integration over the loop momentum and 
therefore can be simply extracted, 
the infrared and collinear singularities appear on the boundaries of Feynman-parameter space.  We must therefore extract the 
infrared and collinear singularities before integrating 
numerically.  To do so we sector-decompose the Feynman-parametric integrals. However, even after 
these divergences are extracted, it is not possible to perform the Feynman-parametric integrations numerically because 
there are singularities 
inside the integration region coming from internal loop thresholds.  These singularities are avoided by deforming 
the integration contour into the complex plane. Once  
infrared and collinear singularities are extracted and the integration 
contour is deformed, we obtain representations of the Feynman parametric 
integrals suitable for numerical integration.  No reduction of tensor integrals is performed.  This allows us to 
bypass many of the issues associated when this algebraic reduction is attempted.

We present below phenomenological results of our NLO QCD calculation of $pp \to t\bar{t}Z$.  We employ the 
Martin-Roberts-Stirling-Thorne parton distribution functions~\cite{mrst} at the appropriate order in the perturbative expansion.  
All numerical results are obtained using the adaptive Monte Carlo integration 
algorithm VEGAS \cite{vegas} as implemented in the CUBA library \cite{cuba}.  We have checked that our results 
satisfy several consistency checks.  The leading order result obtained with our code matches that obtained using the program 
MadEvent~\cite{Maltoni:2002qb}.  Using an eikonal approximation, the divergent parts of the one-loop virtual corrections 
arising from soft gluon exchange can be calculated in a simple analytic form.  We have checked that our full results agree with this form.  
All divergences cancel once we assemble the separate components of the computation discussed above.  The virtual 
corrections are independent of the size of our contour deformation of the Feynman parametric integrals.  Finally, we have implemented 
all parts of our calculation in several independent codes which agree for all observables studied.

\begin{figure}[htb]
\includegraphics[angle=90,width=3.4in]{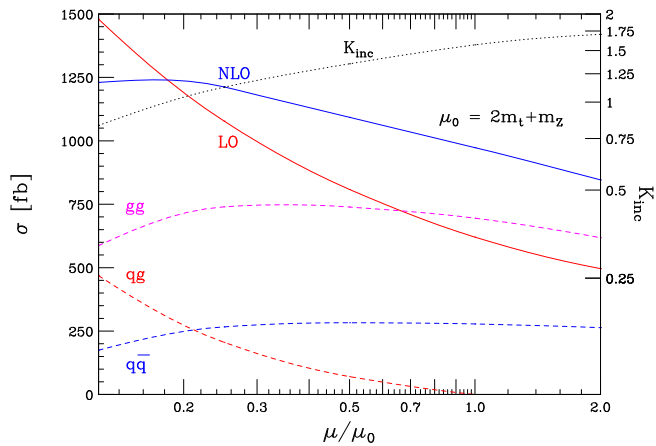}
\caption{Inclusive cross section for $pp \to t\bar{t}Z$ as a function of the scale choice $\mu$.  Included are the LO and NLO results, 
as well as the contributions of the $gg$, $q\bar{q}$, and $qg$ partonic channels at NLO.  The dotted line shows the inclusive $K$-factor; the value 
for this should be read from the axis on the right of the plot.}
\label{incplot}
\end{figure}

We present in Fig.~(\ref{incplot}) the inclusive cross section at both leading order and next-to-leading order in the perturbative expansion.  
We have equated the renormalization and factorization scales to a common 
value $\mu=\mu_R=\mu_F$, and have varied them from $\mu_0/8$ to $2\mu_0$ with $\mu_0=2 m_t+m_Z$.  The dependence of 
the $pp \to t\bar{t}Z$ rate on this unphysical scale parameter is significantly lessened when the NLO corrections are included.  
Choosing a scale too different from the typical momenta and energies that give the dominant contributions to the process leads to 
large logarithms that spoil the convergence 
of the perturbative expansion.  Therefore, to estimate a central value and theoretical error for the cross section, we should a 
pick a scale roughly equal to the typical transverse momenta and masses in the final state and vary $\mu$ around this value.  As these 
momenta and 
masses are 
approximately 100-200 GeV, we consider $\mu$ in $[\mu_0/4,\mu_0]$ a reasonable range of scale variation with $\mu=\mu_0/2$ a good 
central value.  This yields a cross section of 1.09 pb with a theoretical error of $\pm 11\%$ at NLO.  The result at LO is 
0.808 pb with an uncertainty of $\pm 25-35\%$.  The inclusive $K_{inc}$-factor for this process, defined as the ratio of 
the cross section at NLO to that at LO, is $K_{inc}=1.35$ for $\mu=\mu_0/2$.  The variation of $K_{inc}$ with scale is 
also shown in Fig.~(\ref{incplot}); it changes from 1.1 to 1.6 as $\mu$ varies from $\mu_0/4$ to $\mu_0$.  Also included in 
Fig.~(\ref{incplot}) 
are the separate contributions of the $gg$, 
$qg$, and $q\bar{q}$ partonic processes at NLO.  The significant scale dependence of the $qg$ component, which first appears at this order 
in the perturbative expansion, is noteworthy.

\begin{figure}[htb]
\includegraphics[angle=90,width=3.4in]{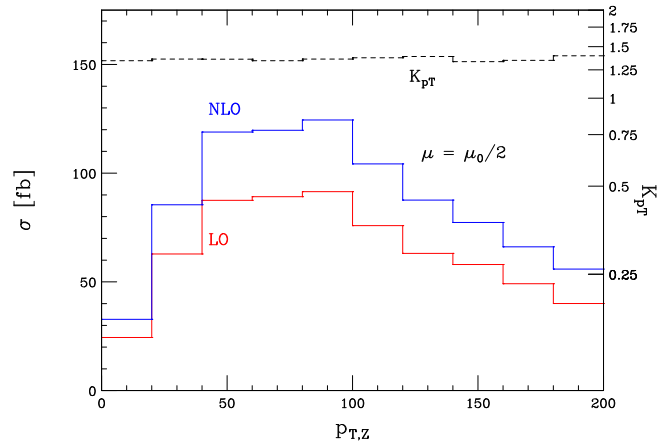}
\caption{Transverse momentum spectrum for $pp \to t\bar{t}Z$ for the scale choice $\mu=\mu_0/2=m_t+m_Z/2$.  Included are the LO and NLO results.  
The $p_T$ dependent $K$-factor for each bin, $K_{p_T}$, is also shown; the value for this should be read from the axis on the right of the plot.}
\label{ptplot}
\end{figure}

In addition to the inclusive cross section, the impact of higher order corrections on differential distributions must be studied.  
An interesting question to consider is whether their effect is completely described by the inclusive $K_{inc}$-factor.  
If so, the NLO corrections can be accurately and simply included in leading order simulation codes by an overall reweighting of event 
rates.  To investigate this question we present in Fig.~(\ref{ptplot}) the bin-integrated transverse momentum spectrum of the $Z$ boson at 
both LO and NLO for the scale choice $\mu=\mu_0/2$.  Most of the cross section comes from events 
with $p_T^Z$ less than 200 GeV.  Included in this plot is the ratio of the NLO $p_T^Z$ distribution over the LO spectrum, 
$K_{p_T}$.  It is flat to within a few percent over the entire range, and is equal to the inclusive value $K_{inc}=1.35$.  The small impact 
of higher order corrections on the $p_T^Z$ distribution can be roughly understood by noting that at tree level, $pp \to t\bar{t}Z$ is 
already a three-body process.  Including additional partonic radiation does not open up new regions of phase space as the $Z$ boson can already 
recoil 
against the $t\bar{t}$ pair.  This intuitive argument leads us to expect that the shape of many other kinematic distributions will also be 
approximately unchanged by NLO corrections.

We can use these results to estimate the improvement in the measurement of $t\bar{t}Z$ couplings at the LHC after NLO corrections 
are included.  Assuming approximate $CP$ conservation, four relevant $t\bar{t}Z$ couplings exist: dimension-four vector and axial couplings, and 
two dimension-five dipole couplings.  Although the measurements of these parameters are correlated, the analysis of Ref.~\cite{ulitop} indicates 
that the dipole couplings are expected to be 
measured with a precision of $\pm 50\%$ at the LHC and $\pm 25\%$ with the super-LHC luminosity upgrade, while the axial coupling 
should be measured with $\pm 15\%$ precision at the LHC and with $\pm 5\%$ at the super-LHC.  This study also found 
that the uncertainty arises almost entirely from the signal normalization 
and statistics; the backgrounds are negligible.  This analysis utilized a scale choice $\mu=m_t$, for which we find 
$K_{inc}\approx 1.3$.  This yields a small improvement of the relative statistical error.  The theoretical uncertainty assumed 
in this analysis of~\cite{ulitop} was $\pm 30\%$.  The authors also studied the expected improvement possible if higher order corrections reduced
this error to $\pm 10\%$, and concluded that improvements in the precisions quoted above could reach a factor of two at the LHC.  A conservative 
estimate of the remaining theory uncertainty from our prediction for $pp \to t\bar{t}Z$ is $\pm 15\%$.  This accounts for imprecise 
knowledge of parton distribution functions.  While the full factor-of-two improvement in the precision of the $t\bar{t}Z$ couplings 
seems slightly out of reach with our current knowledge of the higher order corrections, we still expect a significant improvment 
once this reduced theoretical error is propagated through the full analysis.  Presumably the improvement is more significant with the 
super-LHC luminosity upgrade, since the relative importance of the statistical errors should be decreased.

In summary, we have presented the complete calculation of the next-to-leading order QCD corrections to the 
$pp \to t\bar{t}Z$ scattering process at the LHC.  Study of this channel allows measurements of the $t\bar{t}Z$ electroweak couplings, 
which cannot be directly accessed with current experiments.  These couplings probe 
many forms of physics beyond the Standard Model, such as small mixings of extra heavy gauge bosons or vector-like fermions with 
the Standard Model top quark and $Z$ boson.  To compute these corrections we present a novel approach 
to perturbative calculations in QCD powerful enough to handle the significant complexities that occur when $2 \to 3$ and more 
complicated scattering processes are studied.  This method is highly automated and flexible, and is based on a completely numerical 
algorithm for computing loop integrals.  This is the first time it has been applied to a scattering process that contains the full set of 
complexities that occur in multi-leg one-loop calculations.  
We find that the NLO QCD corrections increase the leading order $pp \to t\bar{t}Z$ cross section by a factor of 1.35 for canonical choices 
of the factorization and renormalization scales.  We estimate the remaining theoretical uncertainty from uncalculated higher order corrections 
to be $\pm 11\%$. Previous studies based on leading order cross sections have found that the normalization uncertainty 
of the $pp \to t\bar{t}Z$ rate is the dominant error in extracting $t\bar{t}Z$ couplings at the LHC.  We estimate that the relative 
precision of the $t\bar{t}Z$ couplings will be improved by a factor of 1.5-2 once our results are incorporated into these analyses.  We find 
that the NLO corrections do not significantly change the shape of the kinematic distributions we have studied, indicating that they 
can be accounted for by an overall scaling of the LO result.

We are excited by the potential of our approach to NLO QCD calculations presented here.  We anticipate and look forward 
to using it to understand other processes of interest at future collider experiments. 

The work of A. L. and K. M. was supported by the US Department of Energy  under contract DE-FG03-94ER-40833.  
The work of T. M. and F. P. was supported by the DOE grant DE-FG02-95ER40896, 
Outstanding  Junior Investigator Award, by the University of Wisconsin Research Committee
with funds provided by the Wisconsin Alumni Research Foundation, and
by the Alfred P.~Sloan Foundation.



\begin{thebibliography}{29}

\bibitem{topd}
F.~Abe {\it et al.}  [CDF Collaboration],
  Phys.\ Rev.\ Lett.\  {\bf 74}, 2626 (1995);
S.~Abachi {\it et al.}  [D0 Collaboration],
  Phys.\ Rev.\ Lett.\  {\bf 74}, 2632 (1995).

\bibitem{ulitop}
U.~Baur, A.~Juste, L.~H.~Orr and D.~Rainwater,
  Phys.\ Rev.\  D {\bf 71}, 054013 (2005);
U.~Baur, A.~Juste, D.~Rainwater and L.~H.~Orr,
  Phys.\ Rev.\  D {\bf 73}, 034016 (2006).

\bibitem{refs1}
A.~Denner and S.~Dittmaier,
  Nucl.\ Phys.\  B {\bf 658}, 175 (2003);
A.~Denner and S.~Dittmaier,
  Nucl.\ Phys.\  B {\bf 734}, 62 (2006);
W.~T.~Giele and E.~W.~N.~Glover,
  JHEP {\bf 0404}, 029 (2004);
R.~K.~Ellis, W.~T.~Giele and G.~Zanderighi,
  Phys.\ Rev.\  D {\bf 72}, 054018 (2005)
  [Erratum-ibid.\  D {\bf 74}, 079902 (2006)];
T.~Binoth, J.~P.~Guillet, G.~Heinrich, E.~Pilon and C.~Schubert,
  JHEP {\bf 0510}, 015 (2005);
C.~Anastasiou and A.~Daleo,
  JHEP {\bf 0610}, 031 (2006);
Z.~Bern, L.~J.~Dixon, D.~C.~Dunbar and D.~A.~Kosower,
  Nucl.\ Phys.\  B {\bf 435}, 59 (1995);
R.~Britto, F.~Cachazo and B.~Feng,
  Nucl.\ Phys.\  B {\bf 715}, 499 (2005);
G.~Ossola, C.~G.~Papadopoulos and R.~Pittau,
  Nucl.\ Phys.\  B {\bf 763}, 147 (2007);
D.~Forde,
  Phys.\ Rev.\  D {\bf 75}, 125019 (2007);
R.~K.~Ellis, W.~T.~Giele and Z.~Kunszt,
  JHEP {\bf 0803}, 003 (2008);
Z.~Bern, L.~J.~Dixon and D.~A.~Kosower,
  Annals Phys.\  {\bf 322}, 1587 (2007).

\bibitem{refs2}
J.~M.~Campbell, R.~K.~Ellis and G.~Zanderighi,
  JHEP {\bf 0610}, 028 (2006);
J.~M.~Campbell, R.~Keith Ellis and G.~Zanderighi,
  JHEP {\bf 0712}, 056 (2007);
S.~Dittmaier, P.~Uwer and S.~Weinzierl,
  Phys.\ Rev.\ Lett.\  {\bf 98}, 262002 (2007);
S.~Dittmaier, S.~Kallweit and P.~Uwer,
  Phys.\ Rev.\ Lett.\  {\bf 100}, 062003 (2008);
V.~Hankele and D.~Zeppenfeld,
  Phys.\ Lett.\  B {\bf 661}, 103 (2008);
  G.~Bozzi, B.~Jager, C.~Oleari and D.~Zeppenfeld,
  Phys.\ Rev.\  D {\bf 75}, 073004 (2007);
T.~Binoth, G.~Ossola, C.~G.~Papadopoulos and R.~Pittau,
  arXiv:0804.0350 [hep-ph].

\bibitem{lmp}
A.~Lazopoulos, K.~Melnikov and F.~Petriello,
  Phys.\ Rev.\  D {\bf 76}, 014001 (2007).

\bibitem{Lazopoulos:2007bv}
  A.~Lazopoulos, K.~Melnikov and F.~J.~Petriello,
  Phys.\ Rev.\  D {\bf 77}, 034021 (2008).

\bibitem{babis1}
C.~Anastasiou, S.~Beerli, A.~Daleo, JHEP {\bf 0705}, 071 (2007).

\bibitem{babis2}
C.~Anastasiou, S.~Beerli, A.~Daleo, arXiv:0803.3065 [hep-ph].

  
\bibitem{Hepp:1966eg}
  K.~Hepp,
  Commun.\ Math.\ Phys.\  {\bf 2}, 301 (1966);
  M.~Roth and A.~Denner,
  Nucl.\ Phys.\  B {\bf 479}, 495 (1996);
  T.~Binoth and G.~Heinrich,
  Nucl.\ Phys.\  B {\bf 585}, 741 (2000).

\bibitem{soper}
D.~E.~Soper,
  Phys.\ Rev.\ Lett.\  {\bf 81}, 2638 (1998);
D.~E.~Soper,
  Phys.\ Rev.\  D {\bf 62}, 014009 (2000);
D.~E.~Soper,
  Phys.\ Rev.\  D {\bf 64}, 034018 (2001);
Z.~Nagy and D.~E.~Soper,
  Phys.\ Rev.\  D {\bf 74}, 093006 (2006);

\bibitem{qgraf} P.~Nogueira, J.~Compt.~Phys. {\bf 105}, 279 (1993).

\bibitem{form}
J.~A.~M.~Vermaseren,
  arXiv:math-ph/0010025.

\bibitem{twocut} 
B.~W.~Harris and J.~F.~Owens,
  Phys.\ Rev.\  D {\bf 65}, 094032 (2002).

\bibitem{mrst} A.D.~Martin, R.G.~Roberts, W.J.~Stirling and R.S.~Thorne, 
Eur. Phys. J. {\bf C23}, 73 (2002); Phys. Lett. {\bf B531}, 216 (2002).

\bibitem{vegas} G.~P.~Lepage, J.~Comp.~Phys.~{\bf 27}, 192 (1978); Cornell University report CLNS-80/447, (1980).

\bibitem{cuba}
T.~Hahn,
  Comput.\ Phys.\ Commun.\  {\bf 168}, 78 (2005).

\bibitem{Maltoni:2002qb}
  F.~Maltoni and T.~Stelzer,
  JHEP {\bf 0302}, 027 (2003).

\end{thebibliography}
\end{document}